\documentclass[a4paper,11pt]{article}
\usepackage{graphicx} 
\usepackage{aaskaiid}
\usepackage{aas_journal_abbr}
\usepackage{orchidlink}

\title{Unravelling the nonthermal emissions from the quiet solar corona with the SKA}
\ShortTitle{Unravelling the nonthermal emissions from the quiet solar corona}

\author[1]{Surajit Mondal\orcidlink{0000-0002-2325-5298}}
\ShortName{Mondal et al.} 
\author[1]{Divya Oberoi \orcidlink{0000-0002-4768-9058}}
\author[2]{Rohit Sharma}
\author[3]{Peijin Zhang \orcidlink{0000-0001-6855-5799}}
\author[4]{Devojyoti Kansabanik\orcidlink{0000-0001-8801-9635}}
\author[5]{Xingyao Chen\orcidlink{0000-0002-1810-6706}}
\author[6]{Eduard Kontar}

\affiliation[1]{National Centre for Radio Astrophysics, Tata Institute of Fundamental Research, Ganeshkhind Road, Pune University Campus, Pune, India, 411007}
\emailAdd{surajit4444mondal@gmail.com}
\affiliation[2]{Indian Institute of Technology, Kanpur, India, 208016}
\affiliation[3]{Centre for Solar Terrestrial Research, New Jersey Institute of Technology, Newark, USA, 19711}
\affiliation[4]{Instituto de Astrofísica de Andalucía-CSIC (IAA-CSIC), Granada, Spain}
\affiliation[5]{State Key Laboratory of Solar Activity and Space Weather, National Space Science Center, Chinese Academy of Sciences, Beijing, China}
\affiliation[6]{School of Physics and Astronomy, University of Glasgow, UK}

\abstract{

Investigation of the nonthermal emissions from the quiet solar corona has been rather limited. This primarily stems from the fact that the emission is expected to be quite weak and very high dynamic range images are required to detect such emissions. Although, the few detections of the nonthermal emissions have all come from the radio band, the past observations had several issues, the most important being the low image fidelity and lack of simultaneous broadband observation capability. Recent observations have been able to tackle many of these prior difficulties using modern instrumentation, and have produced multiple interesting results. In spite of these improvements, multiple challenges remain. For example, most of the recent investigations were only able to focus on the brightest of these events, due to rather poor spectroscopic snapshot PSF of the instruments. The SKA, with its excellent sensitivity as well as its exquisite spectroscopic snapshot PSF would be a game-changer in this field. It would not only allow us to detect and characterise these emissions in Stokes I, but would also allow us to investigate their polarisation properties as well. The high angular resolution offered by the SKA will allow a unique association of the detected radio transients with their thermal counterparts. This would enable us to investigate the thermal-nonthermal energy partition even for these rather weak transient emissions, which in turn would allow validation of particle acceleration and magnetic reconnection theories in a regime vastly different from that done previously. In addition, characterization of the nonthermal emissions of these weak coronal transients may also provide a new probe to understand how energy is transferred from the photosphere and dumped into the corona, and thus serve as a new tool to tackle the coronal heating problem.
}

\begin{document}
\maketitle

\section{Introduction}

The solar corona can roughly be divided into three main regions: active regions, coronal holes and quiet Sun. Active regions are regions with strong closed magnetic field. Coronal holes, on the other hand, are regions of dominant open magnetic field. The easiest way to define the quiet Sun is the region which is not an active region or a coronal hole. All of these regions have their own characteristic emission characteristics at different parts of the electromagnetic spectrum, spanning from the radio to the X-ray wavelengths, and have been studied extensively. Multi-wavelength observations have demonstrated that the solar corona is highly dynamic. Investigating these dynamics is crucial for building a complete understanding of the solar corona. While the thermal characteristics of the dynamic corona have been investigated over a range of temporal and spatial scales, detailed investigations into the nonthermal characteristics have primarily been limited to large explosive phenomena like flares and coronal mass ejections. 
Radio observations are a very powerful tool to probe the nonthermal nature of these transient emissions.
These include the characterization of weak transients, which are important for answering multiple long standing questions in solar physics including addressing coronal heating and solar wind acceleration problem.
Nonetheless, radio investigations of transients have also largely been limited to solar radio bursts and bright flares \citep[see][ for more details]{kumari2026.SKA}. This is because most radio instruments are incapable of delivering data capable enough to produce high dynamic range spectroscopic snapshot images, which are critical to detect and characterize these weak emissions. 

Most investigations into the transient emissions from the quiet solar corona have been conducted in the Extreme Ultraviolet and the X-ray wavelengths. However the EUV and X-ray emission from the higher corona is very faint due to the low electron densities, making these observations primarily suited for investigations of the low corona. Additionally, the bright X-ray and EUV emissions from the low corona and transition region, make it practically impossible to detect the on-disc EUV and X-ray emission from higher coronal heights. Higher coronal heights can only be sampled beyond the limb. Thus detailed investigation about the characteristics of transients in the corona, which spans a very large height range, have been limited. Radio observations, however, would naturally sample different coronal heights, with higher frequencies probing lower in the corona. This is because, due to the increasing density at lower coronal heights, the plasma frequency increases with a decrease in coronal height, resulting in the escape of only higher frequencies from the low corona. Hence radio observations with large bandwidths can probe both the low and high coronal heights. Thus characterizing the transient emissions using wide bandwidth radio observations would provide a more detailed understanding of the solar corona and how the coronal dynamics vary with height.

Multiple theories, including those which try to explain coronal heating and the acceleration of solar wind, allude to continuous magnetic reconnections in the solar corona, including in coronal holes and the quiet Sun regions. While there is a lot of indirect evidence in support of the presence of such small scale magnetic reconnections throughout the solar corona, direct evidence of the magnetic reconnections has remained elusive. Radio observations are excellent probes of magnetic reconnection processes, and have been used to directly measure the evolution of the coronal magnetic field during magnetic reconnections \citep{fleishman2020}. These small scale magnetic reconnections are also expected to accelerate nonthermal electrons, similar to their larger counterparts in flares and CMEs. The accelerated nonthermal electrons, depending on the local physical conditions, emit in the radio wavelengths via different emission mechanisms. Lower in the corona, at GHz frequencies, these electrons primarily emit gyrosynchrotron emission, whereas, at metrewaves, they are believed to emit using the plasma emission mechanism. Thus detecting nonthermal emission at radio wavelengths, can directly suggest the presence of nonthermal electrons. Additionally, since plasma emission is a coherent emission mechanism, its high brightness temperatures make detecting even weak nonthermal electrons less difficult. Observations at GHz frequencies have an added advantage as well. At these frequencies, the dominant emission mechanism is nonthermal gyrosynchrotron emission, which can be modelled to determine both the magnetic field and the nonthermal electron distribution, which can in turn be used to directly constrain the reconnecting magnetic field. 

The nonthermal electron distribution inferred from radio observations can also be used to constrain particle acceleration theories. Particle acceleration theories are often validated against observations of flares and CMEs, which generally are very efficient particle accelerators. However the weak magnetic reconnections occurring in the quiet Sun can have very different magnetic field geometry and plasma conditions than the large explosive events. This would allow investigation of these theories in a very different and unexplored part of phase space, which would ultimately lead to better and more robust theories. 

In Section \ref{sec:review}, we review the existing information about the radio transients from the quiet solar corona at relevant frequencies. In Section \ref{sec:ska}, we discuss how SKA can play a big role in making progress on this front. Finally, in Section \ref{sec:conclusion} we conclude.

\section{Review of past works} \label{sec:review}

The observation frequencies of SKA-Mid and SKA-Low not only probe different coronal heights, but are expected to originate from different emission mechanisms as well. We discuss each of these in the following two subsections.

\subsection{Past investigations at GHz frequencies} \label{subsec:review_ghz_freqs}

\begin{figure}
    \centering
    \includegraphics[trim={0 0 8cm 0},clip,width=0.41\linewidth]{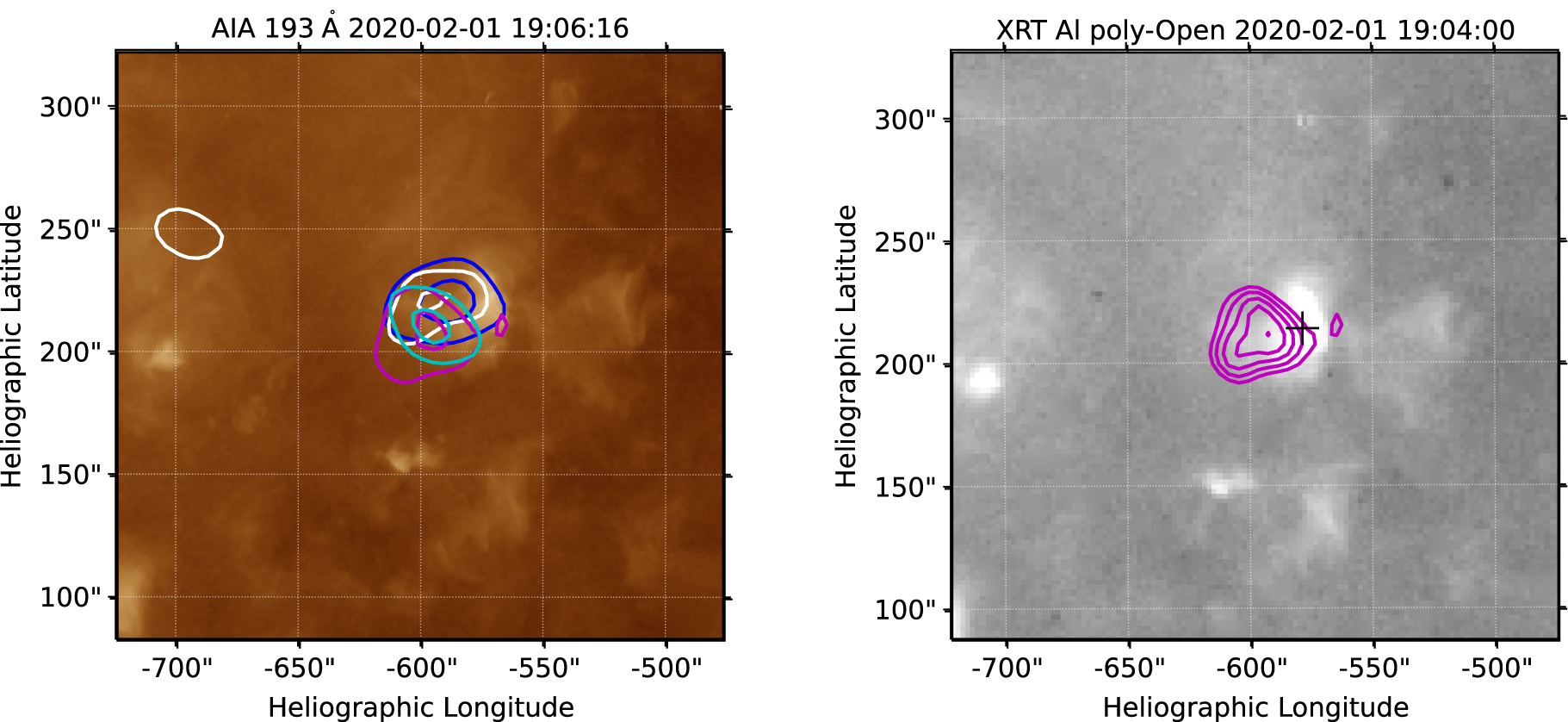}
    \includegraphics[width=0.5\linewidth]{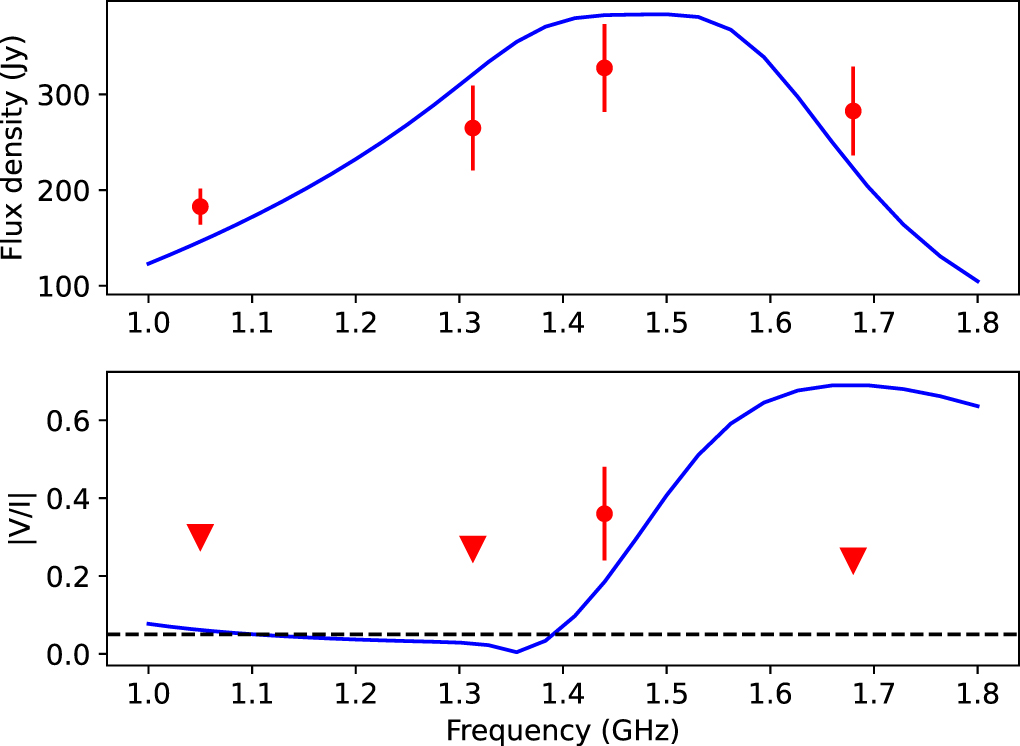}
    \caption{
    The left panel shows the Stokes I radio contours overlaid on top of an EUV image. The blue, white, magenta and cyan contours correspond to images made at 1.057, 1.313, 1.441, and 1.685 GHz respectively. The right panel shows the Stokes I  radio spectrum as well as the variation of the degree of circular polarisation with frequency. The black dashed lines indicates the expected circular polarisation fraction if the emission mechanism is free-free thermal emission.  The blue line is an example of a nonthermal gyrosynchrotron emission spectrum. 
    The figure has been adapted from \citet{mondal2023}.
    }
    \label{fig:mondal2023c_cbp_nonthermal}
\end{figure}

GHz frequencies primarily probe the lower corona and the transition region, the atmospheric layers which have been regularly investigated since the dawn of the space age using EUV and X-ray observations. Combined with this regular EUV and X-ray observations, the availability of sensitive radio instruments like the Very Large Array led to a significant number of investigations into weak transient emissions at these frequencies. However, a lot of these investigations were initially motivated by the observation of variability in soft X-rays outside of active regions. These variable X-ray sources were found to lie over magnetic bipoles, and indicated some potential nonthermal contribution. Such regions are now known as coronal bright points. Keeping in mind, the excellent sensitivity of radio wavelengths to nonthermal emission, efforts were made to investigate coronal bright points at radio wavelengths. However, the early efforts were significantly hindered by constraints on temporal and spatial resolution. \citet{marsh1980}, taking advantage of a solar eclipse, for the first time created very high spatial resolution maps of the Sun and detected radio emission from coronal bright points. The first detection of radio variability associated with coronal bright points was probably reported at 1.4 GHz by \citet{habbal1986}. The authors produced solar maps with a time integration of 2 minutes and demonstrated that the radio sources associated with the coronal bright points were highly variable. They also found that the sources had peak brightness temperatures of about 0.1--0.5 MK, and circular polarization fraction as high as 50\%. The authors hypothesized that the emission is thermal in nature, although they were unable to validate this claim using multi-frequency data. However, under this assumption, they explained the observed circular polarization due to birefringent ray propagation, using which they estimated the magnetic field above the bright points. Similar observations were also carried out at higher frequencies, which resulted in mixed conclusions. \citet{kundu1988} did quasi-simultaneous observation of coronal bright points at 6 cm and 20 cm and found that the spectrum was consistent with free-free emission. However, \citet{krucker1997} demonstrated that the spectrum of at least 5\% of the observed sources were better modelled by a nonthermal gyrosynchrotron model, than by a free-free emission model. Nonthermal emission from transient radio sources away from active regions was also detected by \citet{nindos1999}. However, a key shortcoming of these works was that none of these early investigations were able to do simultaneous observations over a large bandwidth. \citet{mondal2023} probably presented the strongest evidence of nonthermal radio emission from coronal bright points. By making simultaneous measurements over a broad frequency range, they demonstrated that, at least for some cases, the spectrum is inconsistent with a thermal origin. Under the assumption that the emission is powered by nonthermal gyrosynchrotron, they showed that the nonthermal energy associated with the radio transient can be comparable to its thermal counterpart. 
The left panel of Figure \ref{fig:mondal2023c_cbp_nonthermal} shows the Stokes I radio contours overlaid on top of an EUV image. Different colours correspond to the various frequencies at which the same source was simultaneously detected. The right panel shows the Stokes I  radio spectrum as well as the variation of the degree of circular polarisation with frequency. The black dashed lines indicate the expected circular polarisation fraction if the emission mechanism is free-free thermal emission.  The blue line is an example of a nonthermal gyrosynchrotron emission spectrum.  It is evident that the nonthermal emission mechanism is a better descriptor of the observed spectrum.

\subsection{Past investigations at several hundred MHz frequencies} \label{subsec:review_mhz_freqs}

\begin{figure}
    \centering
    \includegraphics[width=0.8\linewidth]{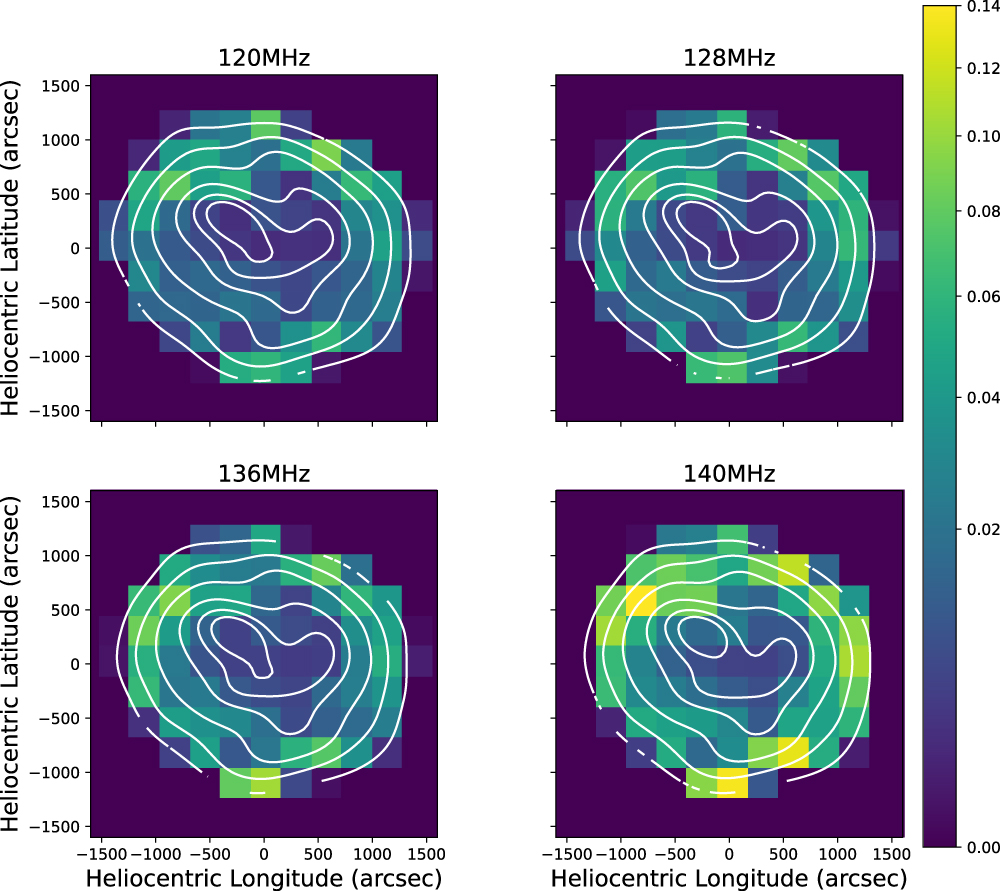}
    \caption{It shows the fraction of observation time for which these transients were detected at any given location of the solar disc. The contours represent the corresponding  Stokes I image, and correspond to 0.2, 0.4, 0.6, 0.8, 0.9, and 0.95 times the peak.
    The figure is adapted from \citet{mondal2023a}. 
    }
    \label{fig:winqse_ubiquitous}
\end{figure}

\begin{figure}
    \centering
    \includegraphics[width=0.8\linewidth]{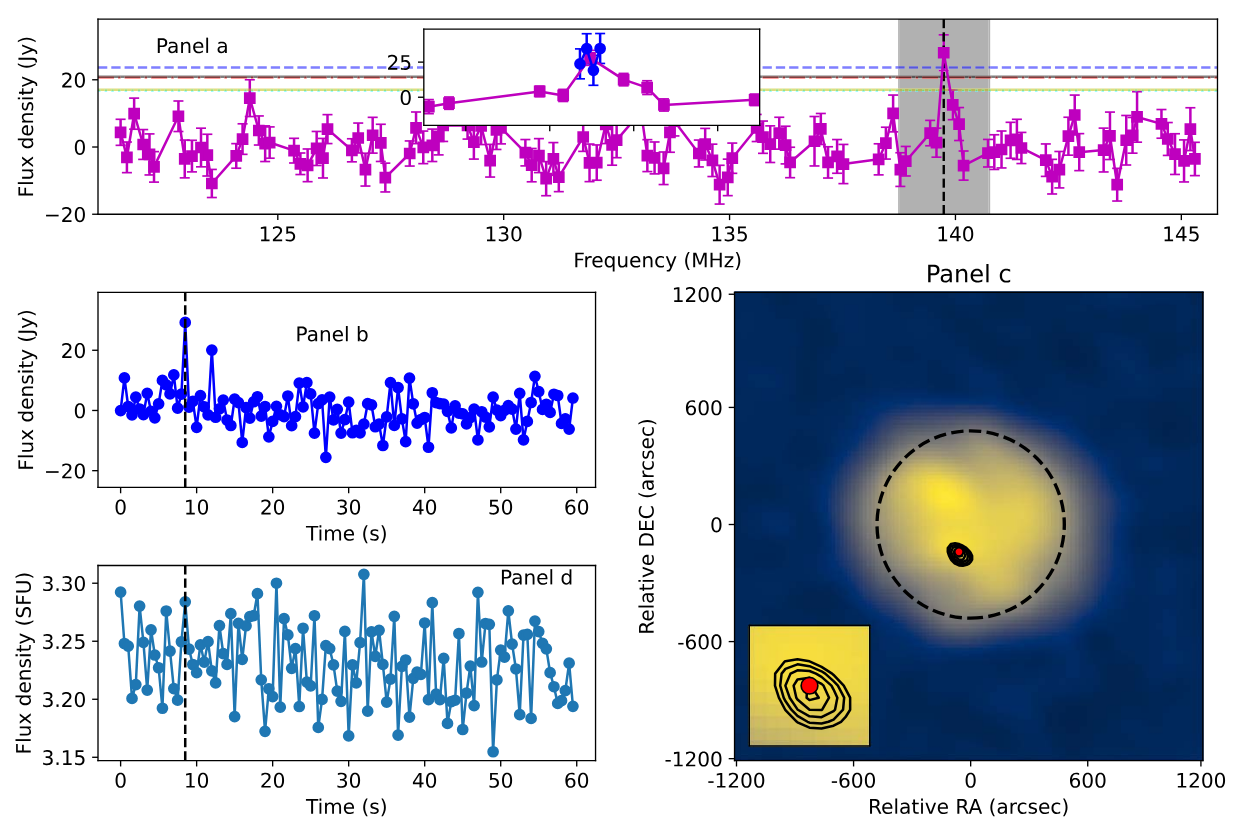}
    \caption{The spectrum and the flux density times series at the spatial location of one example transient are shown in panels (a) and (b), respectively. The magenta and blue points of the spectrum have a resolution of 160 kHz and 40 kHz respectively. 
    Inset in panel (a) shows the zoomed in version of the frequency range indicated by the gray box. These panels clearly demonstrate both the narrowband and impulsive nature of the transient. The image contours corresponding to the peak of the spectrum in panel (a) is shown in panel (c). It clearly shows that no noise peak is present at the flux density level of the WINQSE. Panel (d) shows the variation of integrated flux density of the whole Sun with time at the peak frequency of the WINQSE. The flux density unit in panels (b) and (d) are Jy and SFU (1 SFU=$10^4$ Jy) respectively.
    The figure is adapted from \citet{mondal2023b}. }
    \label{fig:winqse_bandwidth}
\end{figure}

Most of the investigations on small timescale transient emissions at decimetric and metric wavelengths have centered around bright solar radio bursts. Solar radio bursts, like type III, type II, and type I emissions, arise from  coherent emission mechanisms, with brightness temperatures ranging from several MK to hundreds to thousands of MK. All of these bursts are associated with active regions and are not the focus of this chapter. Weak type I radio bursts have also been investigated in multiple occasions \citep[e.g.][]{mercier1997, ramesh2013, Akshay2017}, and are also not relevant here as they were also found to be associated with active regions. \citet{mondal2020} for the first time demonstrated the presence of ubiquitous nonthermal emissions in the quiet solar corona at frequencies spanning 98--169 MHz. They found that the detected nonthermal emissions are very short-lived. Most events had temporal widths of 0.5s, same as the temporal resolution of the instrument. 
Figure \ref{fig:winqse_ubiquitous} shows the fraction of observation time for which these transients were detected at any given location on the solar disc in the sky plane. These data were taken during an exceptionally quiet time during the minima of solar cycle 24. There was no active region on the solar disc during this time. It is evident that these transients were detected throughout the solar disc. \citet{mondal2023b} determined that the bandwidths of these transients is only about a few hundred kHz. \citet{mondal2021} coined the term Weak Impulsive Narrowband Radio Emissions (WINQSEs) to describe these transient emissions. Figure \ref{fig:winqse_bandwidth} shows an image from \citet{mondal2023b} which demonstrates the narrow bandwidth of these transients. Panel b of the figure shows the time series of a region where an example WINQSE was detected. The time of the detection is indicated with a red dashed line. Panel d shows the integrated flux density of the sun. We find that the peak flux density of the WINQSE is only about $30$Jy, whereas the total solar flux density is about 3.2 SFU, were 1SFU=$10^4$ Jy. Thus WINQSEs have a flux density around 0.7\% of the total solar fux density. The authors hypothesized that these are the radio signatures arise from small scale magnetic reconnections happening throughout the solar corona. They suggested that these weak magnetic reconnections give rise to weak nonthermal electron beams, which cannot travel far from their site of origin as they get thermalized very quickly. Examples of fast thermalization of nonthermal electron beam has also been observed in weak Type I bursts \citep{mohan2019}. Presence of WINQSEs have been confirmed using different datasets and techniques as well \citep{sharma2022, bawaji2023, mondal2023a}.

\section{Role of the SKA} \label{sec:ska}

The SKA will be the biggest and the most sensitive radio interferometer\footnote{While solar radio imaging is generally believed to be not limited by sensitivity, high angular resolution solar observations behave in practically the same manner as that of typical astronomical observations. The effect of source noise on a visibility measurement depends inversely on the flux intercepted by the baseline, and hence at very high angular resolution, the source noise \citep{kulkarni1989} becomes negligible.} at its observing frequencies. It will have several new capabilities and would allow a host of different investigations which would otherwise have not been possible. SKA is being built in two parts: the SKA Mid, which operates at the GHz frequencies and the SKA Low, which operates at the metre-wave bands. Considering the different instrument designs as well as the emission mechanisms which are generally observed at these frequency ranges, we discuss the improvements and the pathways to new science that would be opened up by the SKA in the following two subsections.

\subsection{Role of SKA-Mid} \label{subsec:role_ska_mid}

As mentioned in Section \ref{subsec:review_ghz_freqs}, gyrosynchrotron emission is the most commonly observed nonthermal mechanism in the frequency bands probed by the SKA-Mid. While there have been quite a few studies in these bands, all of them suffered from some key limitations:
\begin{enumerate}
    \item All of the past studies used instruments whose spectroscopic snapshot uv-coverage was much sparser than what SKA-Mid will provide. The resultant PSF did not allow high fidelity deconvolution of the quiet Sun, with multiple compact sources of emission present simultaneously. Hence these investigations primarily focused on the brightest of the sources. 
    \item Most prior observations come from instruments which were unable to provide simultaneous wideband coverage. In instances, where wideband data were available, limitations in image fidelity often made it hard to do detailed spectral modelling and infer the energetics of nonthermal electrons.
    \item Poor image fidelity also made it hard to obtain polarization information. 
    \item Most of the early studies were done in an era when simultaneous information about the thermal energetics were limited. Hence detailed characterization of the thermal-nonthermal energy partition for these transients was not possible. 
\end{enumerate}

SKA-Mid will be a game-changer for essentially all of these aspects, as its design offers significant improvements for all of the key aspects limiting progress till now.
 With its array architecture, comprising a dense core and the extended arms, providing long baselines, the SKA will have a very well behaved spectroscopic snapshot PSF, ideal for high fidelity image deconvolution at fine time and frequency resolution. We will no longer be limited to the brightest of these transient sources, but will be able to observe multiple such sources simultaneously. Its high sensitivity will also allow detection of weak circularly polarised flux. If the emission mechanism is gyrosynchrotron, the spectrum can be modelled and the energetics of the nonthermal electrons as well as the magnetic field can be estimated. The high image fidelity expected from the SKA would allow doing such modelling over a large number of such sources, enabling statistical characterization of these weak transient sources. By combining these information, with simultaneous high angular resolution data from EUV and X-rays, it would be possible to determine both the thermal and nonthermal energy associated with these transient sources. Such statistical investigation into the energy partition and magnetic field distribution of these transient sources would lead to new and unique investigations, some examples of which are given below.
\begin{enumerate}
    \item Statistical investigation into the conditions under which nonthermal electrons are detected from the quiet solar corona.
    \item For the first time, we would be able to determine the thermal-nonthermal energy partition of these weak transients, and thus obtain the efficiency with which nonthermal electrons are accelerated in these transients.
    \item This association of the nonthermal gyrosynchrotron emission with other wavebands, would throw a new light on key scientific questions like solar wind acceleration problem. For example, by modelling the nonthermal gyrosynchrotron emission from coronal holes, we would be able to not only measure the energy of the accelerated nonthermal electrons, but also measure the rate of magnetic reconnection, as well as the fraction of magnetic energy which is being used to accelerate the solar wind. This has the potential to revolutionize our understanding of how the solar wind is accelerated to very high speeds.
    \item It is expected that magnetic reconnections occurring in the quiet solar corona have strong guide field. Theoretically, it is expected that in this regime particle acceleration is very inefficient. Similarly, it is expected that in the weak guide field regime, the particle acceleration is highly efficient. However these particle acceleration theories have only been investigated for large explosive events like flares and CMEs, where the reconnection generally happens in the weak guide field regime. Thus statistical investigation of these transients would for the first time allow verification of the magnetic reconnection and particle acceleration theories in the strong guide field regime.
   
\end{enumerate}

\subsection{Role of SKA-Low} \label{subsec:role_ska_low}

As mentioned in Section \ref{subsec:review_mhz_freqs} the work on this front has been rather limited. The flux density of WINQSEs is very low (Fig. \ref{fig:winqse_bandwidth}, less than a few tens of mSFU) \citep{mondal2020, mondal2023a, mondal2023b}, with multiple of them usually being present simultaneously on the solar disc. Hence the image fidelity required to detect WINQSEs is quite high. While such image fidelity and dynamic range was achieved with the MWA, which led to the discovery of ubiquitous nonthermal emissions from the quiet solar corona, detailed characterization was hindered both due to sensitivity and instrument {artefacts} \citep{mondal2023b}. Below we list some key unanswered questions related to these transients, as well as some hypothesis and potential pathways to test them.

\begin{enumerate}
    \item {\bf What is the emission mechanism of WINQSEs?} \citet{mondal2020, sharma2022, mondal2023a, mondal2023b} hypothesised that these transients are produced via plasma emission mechanism. Plasma emission is produced due to nonthermal electrons propagating along magnetic field lines, where the emission frequency is directly dependent on the local plasma density. Thus this hypothesis can be tested by determining the frequency drift of these transients. \citet{sharma2022} presented the first detection of such frequency drift in the spatially resolved dynamic spectrum for one such transient, demonstrating that at least some of these transients are indeed powered by the plasma emission mechanism. However, to do such investigations in a statistical manner, very high dynamic range images at high time and frequency resolution are essential. Additionally, the instrumental {artefacts} should have very low. \citet{mondal2023b} demonstrated that even though MWA was able to detect weak WINQSEs, the instrumental {artefacts} present in the data did not allow accurate bandwidth estimation of these transients, something which is expected to become available with the SKA-Low.
    
    \item {\bf What is the typical energy of the nonthermal electrons responsible for WINQSEs?} While \citet{mondal2023b} suggested that the responsible nonthermal electrons have energies comparable to the thermal energies, they were unable to quantify the energy of the nonthermal electrons. This question can be answered by first determining the responsible emission mechanism and then modelling the observed emission using the determined mechanism. 
    
    \item {\bf What are the signatures of these transients in other wavebands and how is their energy distributed across thermal and nonthermal components?}  The on-disc EUV and X-ray emission are dominated by emissions from low coronal heights, which cannot be probed by SKA-Low. However, off-limb EUV and X-ray observations can provide high angular resolution observations at coronal heights which can also be probed using SKA-Low. To understand the signatures of the radio transients at EUV and X-ray, it is essential to have arcsecond resolution radio observations. After the thermal counterparts are quantified, we need to use the thermal signatures of these emissions to determine the thermal energy, and combine that information with answer of the previous question and determine the thermal-nonthermal energy partition of these transients. 
    
    \item {\bf Are these the radio counterparts of the hypothesised ``nanoflares"?} While \citet{mondal2020, sharma2022, mondal2023a} have demonstrated that these transients are ubiquitous in nature, it is not clear if these are the radio counterparts of the `nanoflares'. This is because the thermal energetics of these transient emissions as well as their importance in coronal heating are unknown. This can only be solved by determining the thermal energetics of these transients in a statistical manner.
    
    \item {\bf How does the occurrence of these transient emissions depend on the coronal height?} The presence of these transients at the metrewave band has already demonstrated that weak magnetic reconnections happen in the quiet solar corona, which in turn produce accelerated nonthermal electrons. However if such weak magnetic reconnections are happening throughout the corona and how they influence the coronal temperature and solar wind are unknown. Answering this question requires high angular resolution radio observations over a wide frequency range.

    \item {\bf How does the occurrence rate of WINQSEs depend on their intensity/energy?}
    For impulsive energy-release processes, a key diagnostic is whether the occurrence rate as a function of intensity or energy follows a power-law distribution, and what the slope of that distribution is. With SKA-Low sensitivity, it will be possible to measure the occurrence rate of WINQSEs down to much lower intensities than is currently feasible and to determine whether their number–intensity (or number–energy) statistics follow a power law. The crucial question is whether the slope is sufficiently steep (i.e. more negative than $\sim\!1$ in number–intensity space, corresponding to a steep index in number–energy space), in which case progressively weaker events contribute an increasingly large fraction of the total energy budget. Establishing this slope robustly is essential for assessing whether WINQSEs can play a significant or even dominant role in quiet-Sun coronal heating.

\end{enumerate}

SKA-Low satisfies all the observational requirements to answer the questions posed above. \citet{mondal2023b} found that the bandwidth of these transients are less than 1 MHz. Most of them are unresolved in time with an instrumental resolution of 0.5 s with the longest ones also being shorter than 10 s \citep{mondal2020, mondal2023a}. The SKA-Low will provide data at much higher time and frequency resolution \footnote{The transient buffer may be useful for this, as it allows for extremely high time and frequency resolution data. While such high resolution data might not be regularly needed, there is a potential to investigate turbulence properties of the solar corona, by characterising the scattering tails potentially visible in WINQSEs dynamic spectrum.}. This along with the very low level of {artefacts} as well as the extreme image fidelity would allow very detailed characterization of these emissions. This would enable detailed theoretical modelling, which, in turn, can then be used to answer many of the questions mentioned above. 

The SKA-Low will also provide a large number of long baselines, which will provide angular resolution comparable to the space-based instruments observing in the EUV and X-ray wavelengths. This would allow us to uniquely associate the detected emissions with their thermal counterparts and thus determine the importance of these weak transients in coronal heating and solar wind acceleration.
Observations of these transients also have the potential of providing high resolution information regarding the turbulent nature of the solar corona. If these emissions are indeed powered by plasma emission, they are expected to be highly sensitive to the coronal scattering. \citet{bawaji2023} made an initial effort to characterize the statistical nature of the morphological properties of these transients, but were unable to relate the observed properties with the turbulent nature of the corona. However, these information can be extracted from the source morphology and location using detailed modelling \citep[e.g.][]{kontar2019, chen_xingyao_2020}. The high angular resolution of the SKA would allow us to determine even the extremely small changes in the source morphology, and thus allow for very fine-grained investigations into the coronal turbulence properties. Additionally, radio observations provide one of the few techniques to measure these key observables as a function of height throughout the solar disc. 

While other wavebands like EUV and X-rays can probe higher heights, those observations are limited to above-the-limb and hence cannot characterize the height dependence of these emissions throughout the solar corona, in contrast to the radio observations. Since the SKA-Low will span a very large frequency range, it will have access to a very large range of coronal heights, making it possible to do these investigations as a function of coronal height.

\section{Discussion and Conclusion} \label{sec:conclusion}

In this chapter, we have reviewed briefly the existing literature on the transient emissions from the quiet solar corona. We have listed some of the key unanswered questions about them and highlighted how the SKA will be instrumental in making progress on them. We have not discussed the calibration and imaging techniques needed to achieve the required image fidelity. We note that at low resolution, both the SKA-Mid and SKA-Low, when observing the Sun would be in the so-called strong source regime {\citep[e.g.][]{bastian2025a, bastian2025b,kulkarni1989,oberoi2026.SKA}.} While the solar observations done till date, have not taken this into account during calibration and deconvolution, we envisage that with this SKA, this will become unavoidable. The calibration accuracy and image fidelity, which we are expecting or demanding from the SKA is very likely to make it essential to take this issue into consideration. With these issues adequately addressed, the SKA would revolutionize investigations on this front, both due to its high sensitivity, high angular resolution, as well as its frequency coverage. The timeline of the SKA is also particularly well matched for these studies. The SKA is expected to be operational by 2030, coinciding with the expected minima of the current solar cycle. This period will be particularly attractive for these studies due to the absence of large active regions and flares associated with very bright transient and persistent radio emissions. Their absence from the solar disk not only substantially reduces the imaging dynamic range requirements it also makes available large quiet Sun regions suitable for these observations, allowing for observations of the quiet Sun for extended periods, making this an ideal period for quiet Sun studies. 

\bibliographystyle{abbrvnat-maxbibnames4}
\bibliography{chapter}
\end{document}